\pdfoutput=1
\documentclass{JINST}

\usepackage{epsfig}
\usepackage{subfigure}
\usepackage{wrapfig}
\usepackage{multirow}
\usepackage{enumerate}
\usepackage{url}

\newcommand{\xphi}{Xeon Phi}

\title{First Evaluation of the CPU, GPGPU and MIC Architectures for Real Time Particle Tracking based on Hough Transform at the LHC}

\author{V. Halyo$^1$\thanks{Corresponding Author}, P. LeGresley$^1$, P. Lujan$^1$, V. Karpusenko$^2$, A. Vladimirov$^2$ \\
$^1$ Princeton University, Princeton, NJ, USA \\
$^2$ Colfax International, Sunnyvale, CA, USA \\
E-mail: \email{vhalyo@gmail.com}}

\abstract{Recent innovations focused around {\em parallel} processing, either through systems containing
multiple processors or processors containing multiple cores, hold great promise for enhancing the performance
of the trigger at the LHC and extending its physics program. The flexibility of the CMS/ATLAS trigger system
allows for easy integration of computational accelerators, such as NVIDIA's Tesla Graphics Processing Unit
(GPU) or Intel's \xphi, in the High Level Trigger. 
These accelerators have the potential to provide faster or more energy efficient event selection, 
thus opening up possibilities for new complex triggers that were not previously feasible.
At the same time, it is crucial to explore the performance limits achievable on the latest 
generation multicore CPUs with the use of the best software optimization methods.
In this article, a new tracking algorithm based on the Hough transform will be
evaluated for the first time on multi-core Intel i7-3770 and Intel Xeon E5-2697v2 CPUs, an NVIDIA Tesla K20c GPU, 
and an Intel \xphi\ 7120 coprocessor. Preliminary time performance will be presented.}


\begin{document}

\section{Introduction} 

Scientific computing is a critical component of the LHC experiment, affecting detector operations, trigger,
simulation, analysis, and the LHC computing grid. The trigger systems are especially sensitive to computing
performance, as they must balance the physics goals of the experiment with the need to be able to process
events in real time. Improvements in the speed with which the trigger system can reconstruct events can thus
extend the physics reach of the LHC. One way this can be achieved is to take advantage of the flexibility of
the trigger system by integrating coprocessors based on Graphics Processing Units (GPUs) or the Many
Integrated Core (MIC) architecture into its server farm.  Parallel processing will provide means not only to
accelerate existing algorithms, but also the opportunity to develop new algorithms that select events that
could have previously evaded detection.

These parallel algorithms will be able to reconstruct in real time all the trajectories of charged particles
in the silicon tracker. It is well-known that the problem of track reconstruction becomes exponentially more 
difficult using traditional reconstruction algorithms as the number of hits in the detector increases due 
to high pileup. The new tracking can not only overcome the challenges faced by the trigger at high pileup, but it 
allows us to develop complex triggers that could select in real time final state topologies not possible before.

Leveraging processors derived from consumer products helps minimize costs associated with purchasing and
operating large computing installations, such as those required for the LHC, but there is also a software
aspect that needs to be considered.  Parallel processors, in the form of multi-core CPUs and, more recently,
highly programmable Graphics Processing Units (GPUs), may require a significant rethinking of algorithms and
their implementations.  At the same time, the High Energy Physics (HEP) community is at a crossroads with
tentative confirmation of the Higgs particle completing the search for particles predicted by the Standard
Model. The search for beyond the Standard Model (BSM) physics will require development of advanced algorithms
for detecting rare new physics phenomena, and these new algorithms will need to be designed and implemented
based on knowledge of the current state and likely future directions for processor architecture.

In this article, a new tracking algorithm based on the Hough transform will be evaluated for the first time on
an Intel \xphi\ and multi-core Xeon CPUs, and compared with performance on an NVIDIA Tesla K20c
GPU. Preliminary time performance will be presented.

\section{Physics Motivation}

Various BSM extensions predict the existence of new, strongly interacting particles that lead to final states
with high jet multiplicities~\cite{bib:sixjets,bib:multijets,bib:tcgut,bib:exoquarks,bib:unichiral,bib:altchiral}.  
Other exotic models might include boosted jets~\cite{Thaler:2008ju,Altheimer:2012mn}, R-parity-violating SUSY models
with long-lived neutralinos~\cite{Carpenter:2007zz,Graham:2012th}, long-lived neutral particles decaying at
macroscopic distances from the primary vertex~\cite{Strassler:2006ri,Strassler:2006im,Halyo:2013yfa} in the
tracker, quirks (also known as iquarks)~\cite{Cheung:2008ke}, displaced black holes~\cite{Halyo:2013cza}, or
long-lived Higgs particles~\cite{Strassler:2006ri,Halyo:2013yfa}. The key to observing these events\cite{Halyo:2013iba}
is a fast tracking algorithm executed in real time in the trigger system that allows us to reconstruct all
prompt (i.e., tracks that originate from the interaction point) and non-prompt tracks for further careful, offline analysis.

Both ATLAS~\cite{Aad:2008zzm} and CMS~\cite{Chatrchyan:2008aa} use a standard multi-level trigger system, with
the lowest level (L1) using fast and simple criteria implemented in hardware and firmware, and higher-level
triggers which reconstruct the full event in software and can apply selection criteria similar to those used
offline, using a farm of commercial CPU processors. This selection is designed to select only the events which
contain interesting physics processes. However, the need to process the events in real time imposes very
stringent limits on the complexity of the analysis that can be done at the trigger level. For instance, in the
offline reconstruction at CMS, tracks can be reconstructed even if they originate relatively far away from the
primary interaction point. However, reconstructing these tracks is computationally expensive, as the number of
possible combinations is quite large; hence, displaced tracks are not reconstructed in the CMS high-level
trigger (HLT) and so one cannot directly trigger on events which contain such a signature.

The flexibility of the trigger system makes it amenable to adding new kinds of hardware, such as GPU- or
MIC-based coprocessors. Thus it is easy to integrate the hardware necessary for the development of massively
parallel algorithms that will not only reconstruct in real time all the particle trajectories in the event,
but allow the development of new algorithms that can select final state topologies previously not possible in
the trigger. This article will evaluate the most computationally expensive application of the tracking
algorithm for the first time on various hardware architectures. These benchmarks are critical for any future
LHC trigger upgrade.

\section{Processor Architecture}

In the past decade, it has become apparent that it is no longer possible to rely solely on increases in
processor clock speed as a means for extracting additional computational power from existing architectures.
The underlying reasons are complex, but center around reaching what may be fundamental limitations in
semiconductor device physics.  For this reason, recent innovations have focused around {\em parallel}
processing, either through systems containing multiple processors, processors containing multiple cores, and
so on.

\subsection{Multi-core CPUs}

Traditional CPUs have been evolving in recent years mainly through the growth of parallelism rather than
increasing clock frequencies. Multiple cores on a single chip have been introduced; the Non-Uniform Memory
Access (NUMA) architecture was incorporated, which allows transparent use of multiple CPU sockets within a
single shared memory system; and vector processing units have evolved to new instruction sets and longer
vector registers. At the same time, the architecture of CPUs has become more efficient with deeper pipelines,
smarter branch prediction and hardware prefetching of data from memory into caches.

Generally speaking, the CPU architecture is resource-rich and is capable of running workloads with complex and
non-local memory access patterns, with or without task and data parallelism. However, when data locality, task
parallelism (multi-threading) and data parallelism (vectorization) are present in the application, the
potential performance benefits become tremendous. For instance, the AVX extensions (Advanced Vector
Extensions) available in modern CPUs offer the ability to quickly operate on floating-point
vectors. Optimization for such CPU systems poses considerable challenges; however, it also promises
significant benefits.

\subsection{Graphics Processing Units}

One interesting technology which has continued to see exponential growth is graphics processing.  A modern
Graphics Processing Unit (GPU) is a massively parallel processor with thousands of execution units to handle
highly parallel workloads related to computer graphics.  By making these execution units highly programmable,
manufacturers have made the massive computational power of a modern GPU available for more general-purpose
computing, as opposed to being hard-wired for specific graphical operations.  In certain applications that can
be executed in massively parallel fashion, this can yield several orders of magnitude better performance than
a conventional CPU. Manufacturers have taken advantage of these possibilities to release GPUs designed for
general-purpose computing, such as the NVIDIA Tesla line. To facilitate the use of general-purpose GPU
computing, NVIDIA has also developed CUDA (``Compute Unified Device Architecture'')~\cite{bib:CUDA}, which
allows rapid development of GPU software in nearly standard C code.

\subsection{MIC architecture (Intel \xphi\ coprocessors)}

With the growing success of GPUs as a massively parallel processor well suited to more general purpose applications, 
Intel has introduced a line of products based on a Many Integrated Core (MIC) architecture, marketed under the name 
Xeon Phi. The \xphi\ coprocessors are symmetric multiprocessors; they physically look similar to a GPU in that they 
plug into a host system via PCI Express. They run a $\mu$OS Linux Operating System. However, a Xeon Phi coprocessor
cannot be used as a stand-alone processor, and requires a host system to operate. 

From an architectural perspective, they also have many similarities to GPUs. Xeon Phi is based on an x86
Pentium core architecture from the early 1990s.  This is a much simplified, and hence smaller, core compared
to modern x86 CPUs.  To make these simplified cores more computationally powerful, 512-bit-wide vector units
have been added to the core.  Because of the simplified nature of the core each coprocessor features 60+ cores
clocked at 1 GHz or more, supporting 64-bit x86 instructions. The exact number of cores depends on the model
and the generation of the product. These in-order cores support four-way hyper-threading, resulting in more
than 240 logical cores. The cores of an \xphi\ coprocessor are interconnected by a high-speed bidirectional
ring, which unites the L2 caches of the cores into a large coherent aggregate cache over 30~MB in size.  The
coprocessor is equipped with 6 to 16~GB of on-board GDDR5 memory.  The speed and energy efficiency of \xphi\
coprocessors come from their vector units.  Each core contains a vector processing unit (VPU) with 512-bit
SIMD (Single Instruction Multiple Data) vectors supporting a new instruction set called Intel Initial
Many-Core Instructions (Intel IMCI).  The Intel IMCI include, among other instructions the Fused Multiply-Add
(FMA), reciprocal, square root, power and exponential function operations, commonly used in physical modeling
and statistical analysis.  The theoretical peak performance of an \xphi\ coprocessor is 1~TFLOP/s in double
precision.  This performance is achieved at the same power consumption as in two Xeon CPU processors, which
yield up to 2-3 times less GFLOP/s.

The greatest difference between the MIC architecture and GPUs is that it is possible to develop and optimize a
single code in C, C++ or Fortran to use on both a multi-core CPU and on a Xeon Phi coprocessor. In contrast,
GPU applications based on the CUDA architecture are built from a code that is different from the CPU
application code, in terms of both syntax and algorithm. This programming model continuity of the MIC
architecture is attractive when the application is developed not just for the accelerator, but for a
heterogeneous system composed of CPUs or coprocessors, or when the application is deployed to clients which
may or may not have access to a coprocessor.

\section{Hough Transform Algorithm}

As a demonstration of how computationally intensive 2D tracking algorithms can take advantage of massively
parallel GPU processing, the authors developed an implementation of the Hough transform using
CUDA~\cite{Halyo:2013iba} in order to measure in real time the transverse momentum of prompt or non-prompt
tracks.  The Hough transform~\cite{bib:HT1,bib:HT2,bib:HT3} is an image processing algorithm for feature
detection that considers all possible instances of a parameterized feature such as a line or circle.  Each
possible instance of a feature starts with zero votes in the parameter space, and then for each piece of input
data votes are added to the feature instances that would include that input data.  After all input data has
been processed the votes in the parameter space are processed.  Locations in the parameter space with more
votes are likely to be actual features in the input data so this step amounts to looking for local maxima in
the parameter space.  Once candidate features have been identified, more expensive computations can be applied
to confirm the existence of the feature.  One should note that the Hough transform approach as presented above
is computationally expensive. Hence, simplification of the standard Hough transform
method~\cite{Ohlsson:1991eh,bib:AHT} had to be used in the past for the purpose of track finding in high
energy physics~\cite{Bailey:2004ga,bib:cosmicmu}.

In practice, the Hough transform is implemented using a discrete grid in the parameter space. In our case, the
parameter space consists of two variables --- the track curvature $\rho$ and the azimuthal angle $\phi$ -- and
the parameter space is discretized into a 2048x2048 grid, which offers a balance between calculation speed and
precision. The first step then consists of finding and incrementing the appropriate grid spaces (voxels) in
parameter space for a given hit, and the second step consists of searching for local maxima on the grid.

\begin{figure}[!Hhtb]
\begin{center}
  \subfigure[]{\includegraphics[width=0.32\textwidth]{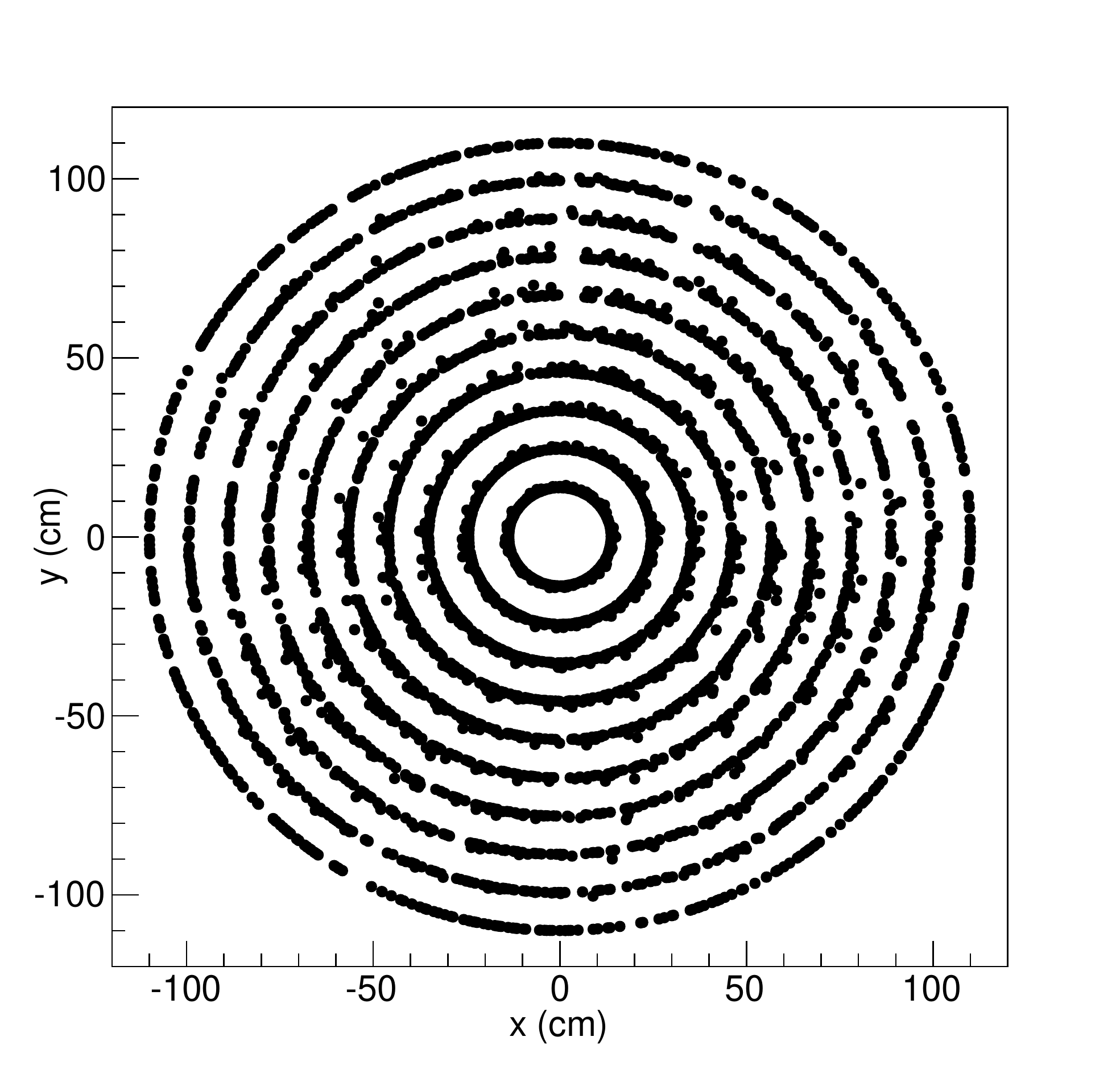} \label{fig:hits}}
  \subfigure[]{\includegraphics[width=0.30\textwidth]{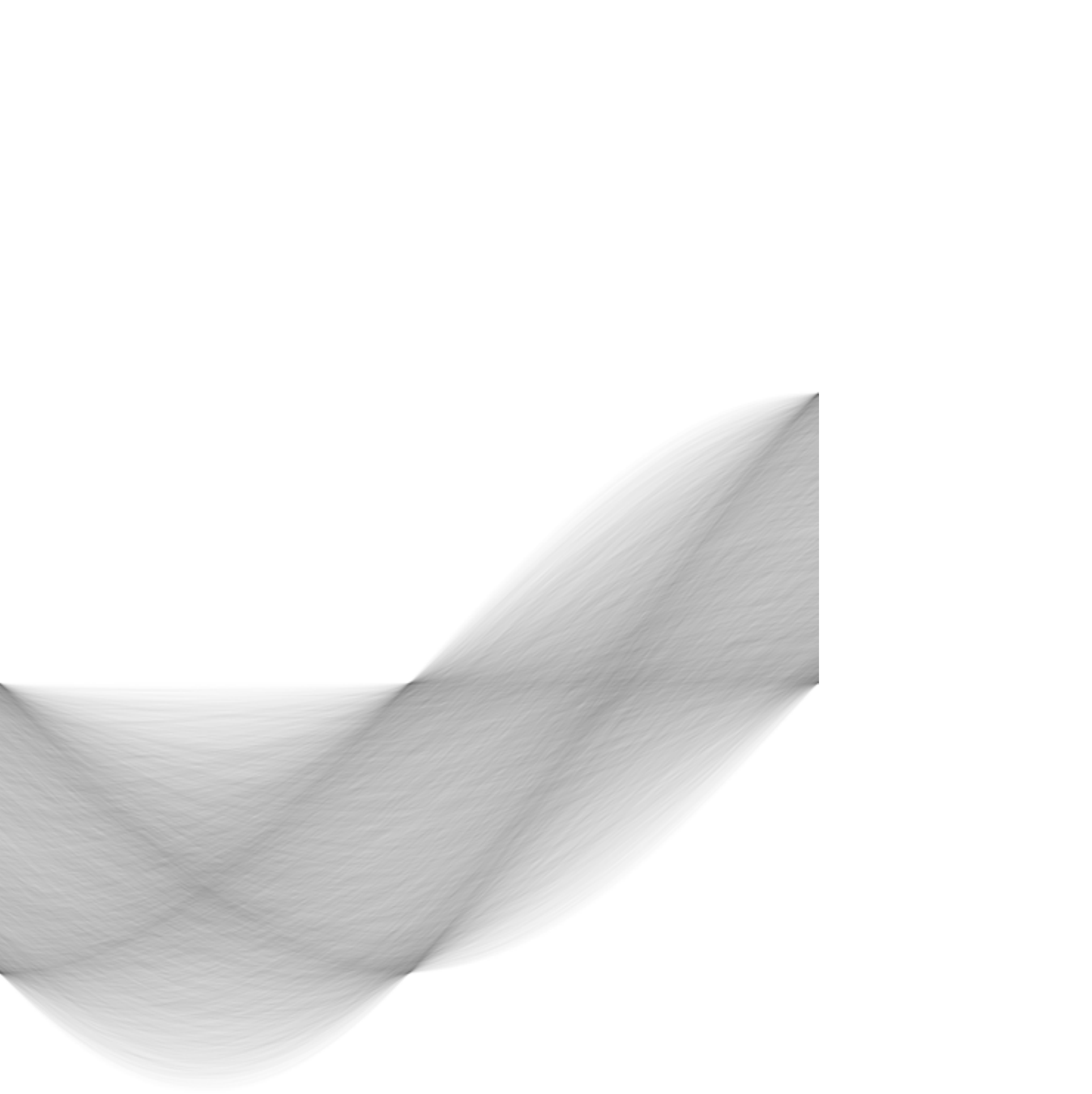} \label{fig:accumulator}} 
  \subfigure[]{\includegraphics[width=0.32\textwidth]{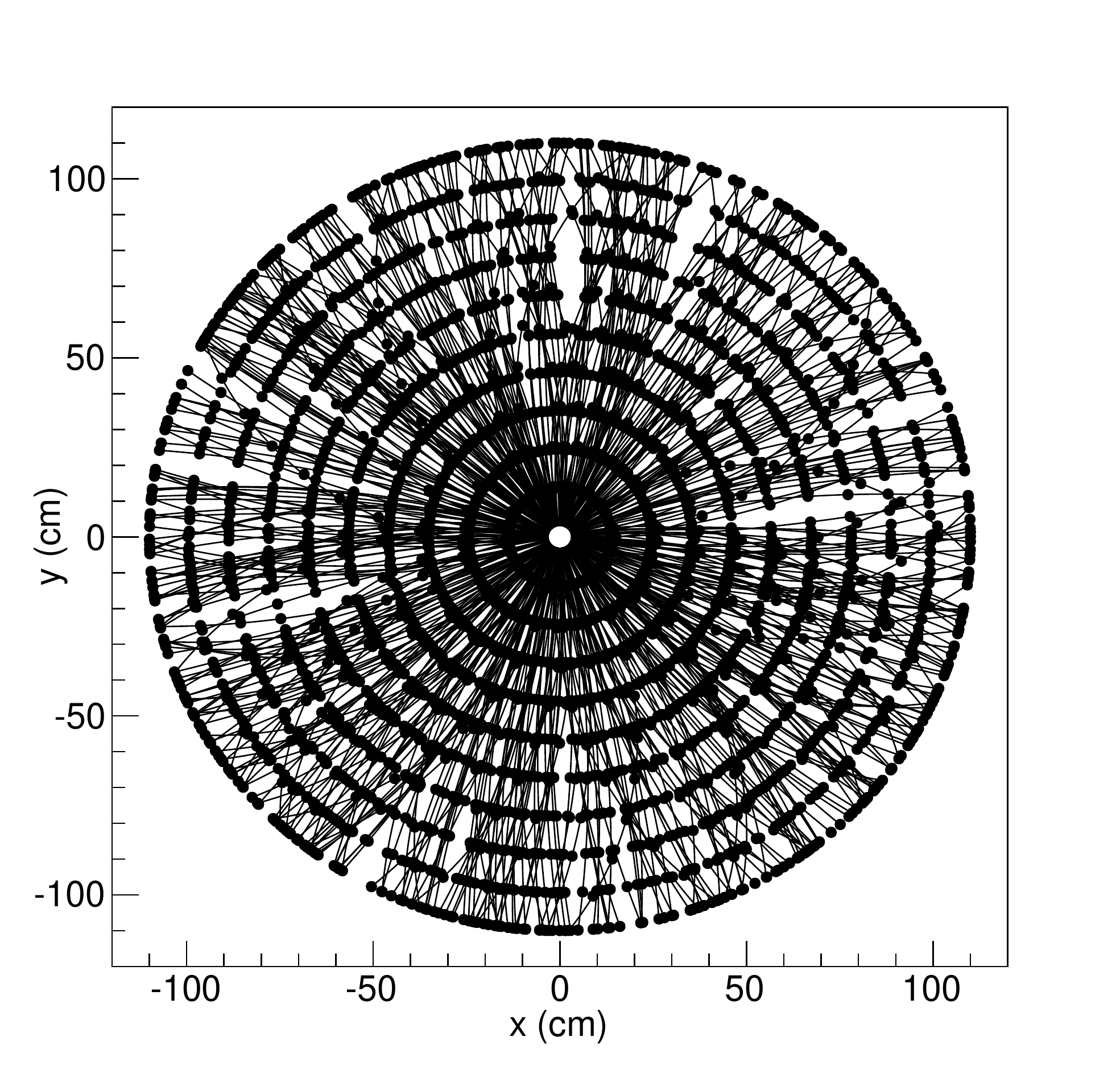} \label{fig:tracks}}
  \caption{Hough transform algorithm applied to a simple example. Left: Hits in a simulated event with curved tracks. 
          Centre: Each hit results in a curve of votes in parameter space. Locations with many votes are likely to be
           tracks in the original data. Right: Candidate tracks identified from finding local maxima in the parameter 
           space.\label{fig:hough}}
\end{center}
\end{figure}

Figure~\ref{fig:hough} shows the algorithm in operation for a simple case with 500 curved tracks.  The
simulated curved track data shown in Figure~\ref{fig:hits}, can correspond to many different possible tracks
in parameter space as shown in Figure~\ref{fig:accumulator}.  When integrated over the entire data set, peaks
appear in parameter space at the values corresponding to the actual, physical trajectories as shown in
Figure~\ref{fig:tracks}.

One clear advantage of the Hough transform over the combinatorial track finding algorithms that are currently
standard is that the execution time is linear with respect to the number of hits present in the event, while
combinatorial algorithms, as the number of combinations increases much more rapidly with respect to the number
of hits, show a worse dependence on the number of hits. In addition, the Hough transform is naturally tolerant
of missing hits or hits that do not exactly fit the candidate features, due to limited resolution or the
discretization used in computing the parameter space. While the Hough transform is more expensive to implement
for a small number of hits, the ability to take advantage of multithreading offers the chance of significantly
improved performance for events with a large number of tracks.

One should note that this paper does not imply that executing machine vision and pattern recognition
algorithms is necessarily more appropriate for parallel processing compared to the conventional combinatorial
algorithms used in high energy physics. Rather, it is a complementary way that the authors began studying due
to the similarity of the problem with computer image analysis. These techniques have proven successful in
image processing using the Hough transform on GPUs.

In the following studies, the sample input data was generated using a Monte Carlo simulation of a simple
detector model where only the transverse plane is considered.  The model contains a simulated beam pipe with a
radius of 3.0 cm surrounded by ten concentric, evenly-spaced tracking layers with an overall radius of 110.0
cm.  A hit resolution of 0.4 mm in each direction is used, corresponding to a relative $p_T$ resolution of
approximately $7\%$ at 100~GeV$/c$.

In previous studies, we have presented preliminary results using a GPU implementation of the Hough
transform~\cite{Halyo:2013cza,Halyo:2013iba}. This paper shows the first results of an implementation on the
Xeon CPU and Xeon Phi architecture, along with improvements in the GPU implementation.

\section{System Description}

\subsection{CPU Architecture}

The CPU performance for the Intel Xeon processor is measured on a Colfax CX2265i-XP5 server~\cite{cx2265i} with
128~GB of 1600~MHz DDR3 ECC memory, based on a two-socket Intel Xeon E5-2697v2 (Ivy Bridge) CPU for servers.
The thermal design power (TDP) of each CPU socket is 130~W.  The sockets are interconnected by a Quick Path
Interconnect (QPI) link and forming a shared-memory NUMA system.  Each socket has 12 physical cores (24 cores
in the system) clocked at 2.7~GHz (turbo frequency of 3.5~GHz) with two-way hyper-threading.  The vector units
of the system support the AVX instruction set with 256-bit vector registers.

The host operating system is CentOS 6.4 Linux with kernel version 2.6.32-358.11.1.el6.x86\_64. The code was
compiled with the Intel C++ compiler version 13.1.3.

The CPU performance for the quad-core Intel i7-3770 processor is measured on a different desktop system, which
is also the host of the NVIDIA Tesla GPU. The TDP of the i7-3770 CPU is 77~W.

\subsection{NVIDIA GPU Architecture}

The NVIDIA Tesla GPU model used for the results presented in this paper is the Tesla K20c (active-cooled model
for workstations), which contains 2496 cores clocked at 706 MHz, 5 GB of on-board memory with a clock of 2.6
GHz, and a 320-bit memory interface.
The code was written in CUDA C and compiled with the nvcc compiler, version 5. The host operating system is
CentOS 6.4 Linux with kernel version 2.6.32-358.11.1.el6.x86\_64.

\subsection{Intel Coprocessor Architecture}

The Xeon Phi coprocessor performance is measured on the same system as the CPU performance.  The system
contains one \xphi\ coprocessor of the QS-7120P series (passive-cooled model for servers) with 61 cores at
1.33~MHz, C0 coprocessor stepping, and 16~GB of GDDR5 RAM at 2750~MHz. The driver stack is MPSS version
2.1.6720-13.

\section{Optimization}

A time profile of the tracking algorithm shows that 92\% of the algorithm execution time is spent on executing
the Hough transform itself. Therefore, we optimize the Hough transform separately on three different
architectures. 

\subsection{Optimization for Tesla GPU}

Extensive work had already been done to implement and optimize the Hough transform for the NVIDIA GPU
architecture~\cite{Halyo:2013cza},\cite{Halyo:2013iba}. These past optimizations included minimizing expensive
trigonometric functions, developing an efficient memory access pattern for reading and writing global memory,
and safely handling updates of values in the parameter space to avoid race conditions.  Additional performance
optimizations have been undertaken to further improve performance, including:

\begin{itemize}
\item additional reductions in global memory accesses;
\item an additional optimization of the global memory access pattern to maximize efficiency of memory coalescing;
\item replacement of any remaining atomic memory accesses to global memory with atomic memory accesses to shared memory.
\end{itemize}

The global memory optimizations are particularly important because it is implemented as off chip
Dynamic Random Access Memory (DRAM) and is much more costly to access (in terms of both latency and
bandwidth) compared to on chip memory locations.  However, the use of global memory is unavoidable
because it is the only memory location available for storage of data that needs to persist over a sequence of
computational operations on the GPU.

\subsection{Optimization for Intel i7 and Xeon CPUs and for Intel Xeon Phi Coprocessors}

Although Hough transform implementations exist in standard libraries such as the Intel Performance Primitives
(IPP) library, these do not take advantage of multi-threading.  For that reason, in this work, we created our
own parallel implementations and optimized them to extract the best performance out of the parallel compute
devices that we benchmarked.

Because of the similarities of Intel i7 and Intel Xeon CPUs with \xphi\ coprocessors, we developed a single
code for all three platforms.  The executables for CPUs and coprocessors are not binary-compatible, and the
C++ code must be compiled twice: once for the CPU and another time for the coprocessor.  We used preprocessor
macros to insert different values of tuning parameters into the CPU and the MIC architecture binaries.

Other than the tuning parameters mentioned above, the optimization methods for both platforms are common:

\begin{itemize}
\item thread parallelism was implemented in the OpenMP framework~\cite{openmp} for the outer loops of the Hough transform
and local maxima search;
\item where possible, the code was written in such a way that the compiler is able to perform automatic vectorization;
\item cross-thread synchronization was avoided by the use of the OpenMP reduction facility and thread-private data storage;
\item where synchronization was necessary (e.g., for merging thread-private storage into a shared storage
array), improved synchronization methods were used, such as ordered for-loops in OpenMP;
\item strip-mining and blocking techniques (see, e.g., \cite{tiling} and references therein) were applied to
nested loops in order to improve the data access locality. These techniques change the order of operations in
nested loops, so that data loaded into caches is re-used as soon as possible, which reduces the frequency of
long-latency memory accesses;
\item in order to utilize the coprocessor, data was moved from the CPU to the coprocessor using the offload
functionality provided by the Intel C++ compiler;
\item data persistence on the coprocessor was used to avoid a reallocation penalty when multiple frames are
offloaded and analyzed.
\end{itemize}

Thanks to the NUMA architecture, a two-socket system acts as a single CPU from the programming perspective,
with OpenMP threads seamlessly scaling across both sockets. Therefore, no additional development was required
to utilize two CPU sockets.  However, in order to bind software threads to the respective core caches, we
enforced thread affinity, which prevented software thread migration across logical cores.  This was done by
setting the environment variable \mbox{\texttt{KMP\_AFFINITY=compact,granularity=fine}} prior to running the
benchmarks.

\section{Preliminary Performance Results}

\begin{figure}[!Hhtb]
\begin{center}
\includegraphics[width=0.75\textwidth]{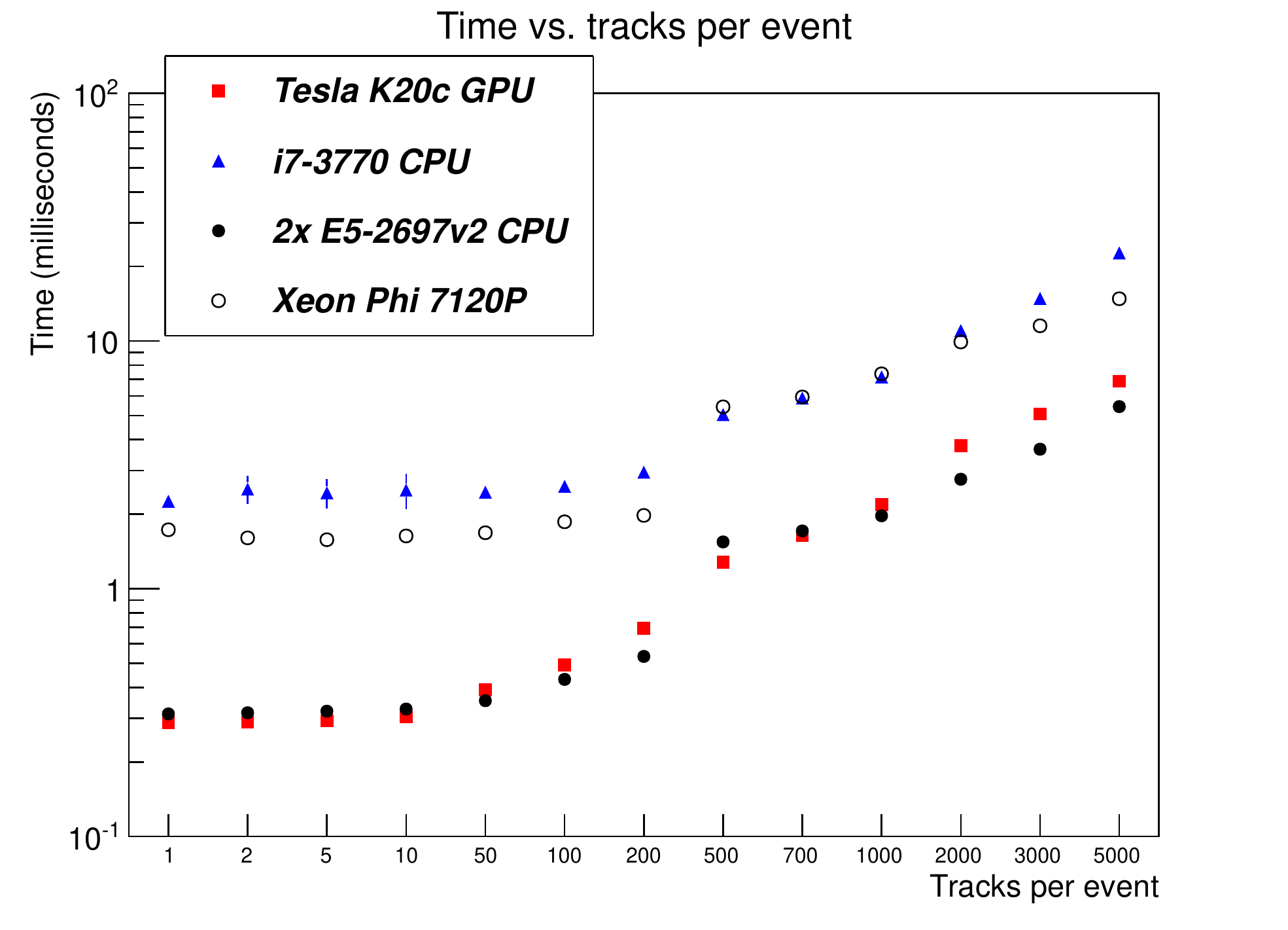} 
\caption{Performance of the Hough transform algorithm on four platforms: NVIDIA Tesla K20c GPU (red squares),
Intel i7-3770 CPU (blue triangles), dual-socket Intel Xeon E5-2697v2 CPU (solid black circles), and Intel Xeon
Phi 7120P coprocessor (open black circles), as a function of the number of simulated tracks in the event. A
2048x2048 grid is used in the parameter space. The error bars indicate variations between successive runs,
which are negligibly small in most cases.\label{fig:TimePerformance}}
\end{center}
\end{figure}

Figure~\ref{fig:TimePerformance} shows the resulting time performance as a function of the number of tracks in
the event for four computing platforms: an Intel i7-3770 CPU, a dual-socket Intel Xeon E5-2697v2 CPU system,
an NVIDIA Tesla K20c GPU, and an Intel Xeon Phi 7120P coprocessor.  The reason for comparing the accelerators
to a dual-socket CPU system (rather than single-socket) is that the TDP of the GPU and of the coprocessor is
225 and 300~W, respectively, whereas a single CPU chip is rated at 130~W.  Therefore, two CPUs make for a more
meaningful watt-to-watt comparison.

According to Figure~\ref{fig:TimePerformance}, the GPU performs almost as fast as dual Xeon CPUs for all event
sizes. The Xeon Phi coprocessor and the i7 CPU are considerably slower. For large events (5000 tracks), the
Xeon Phi is 3x slower, and the i7 is 5x slower than the Xeon system. The 5x ratio of the Xeon to i7
performance is equal to the ratio of the numbers of cores in these platforms corrected for the difference in
their clock frequency. For small events (fewer than 200 tracks), the processing time reaches a plateau at a
level that is is almost an order of magnitude greater for Xeon Phi and i7 than for Xeon and the GPU.

\section{Discussion}

The Hough transform calculation is a highly parallel task. However, it cannot take advantage of the GPU or
many-core architecture in an optimal way because of the nature of this calculation.

The first part of the calculation counts the number of points in the voxels of $(\rho, \theta)$ space.  This
operation has indirect memory accesses in the form \texttt{A[B[i]]+=1}, i.e., the data are written to an array
index, which itself is looked up from another array.  Due to a stochastic pattern of memory access, this
operation cannot be performed with streaming memory accesses, which is a sub-optimal operation regime for the
GPU and the MIC architecture.  On the GPU this is mitigated by replacing accesses to global memory with
accesses to on-chip shared memory.  For the Xeon Phi coprocessor, in the absence of hardware prefetching for
the Level~1 cache, performance in cases like this is dependent on software prefetching. Future versions of the
Intel C++ compiler may be able to automatically implement software prefetching for indirect array accesses and
improve the performance \cite{rk2013}.

The second part of the calculation, where the transformed data set is searched for lines, has a more regular
memory access pattern; however, a considerable part of this calculation uses scalar instructions and a
non-trivial reduction pattern.  This is also a sub-optimal workload for the GPU and the MIC architecture.

GPU and Xeon Phi architecture are designed as computing accelerators and have a higher theoretical peak memory
bandwidth and arithmetic performance than the Xeon or i7 CPU. However, for this application, they do not
provide significant acceleration compared to the multi-core Xeon processor.  In fact, the Xeon Phi coprocessor
is even falling behind in performance behind the Xeon processor of comparable TDP by a significant factor.
This result is expected, because as mentioned above, the nature of the problem does not allow the GPU and the
Xeon Phi coprocessor to efficiently exploit their architecture. At the same time, the Xeon CPU has a
resource-rich architecture, which compensates for sub-optimal memory and operations traffic with its large
unified Level 2 cache, hardware prefetchers for the Level 1 and Level 2 caches, higher clock frequencies of
the cores, and shorter vector units.

Nevertheless, even at the currently observed performance, the inclusion of a GPU or a Xeon Phi coprocessor
into the system configuration may be justified in some cases:

\begin{enumerate}[a)]
\item If the calculation is offloaded to the accelerator (GPU or coprocessor), 
the host CPU is free to do other tasks, such as preparation or post-processing
of data.

\item When it is desirable to achieve the best possible performance within a single chassis,
 or within a limited rack space in a cluster, it is possible to use compute nodes with up to eight accelerators
 (GPUs or Xeon Phi coprocessors) such as \cite{cx9100},\cite{AMAX}.
In addition, a Xeon CPU may be used simultaneously with accelerators to process
a part of the workload. 
Programming models for the CUDA and the Xeon Phi architecture allow easy
implementation of heterogeneous work sharing between CPUs and multiple accelerators (see, e.g., \cite{colfax2013} and \cite{wilt2013}).

\item Finally, in cases where a cluster of compute nodes is deployed for solving the problem of track
detection in parallel, it may be advantageous to ramp up performance by equipping the compute nodes with
accelerators instead of increasing the number of compute nodes.  The accelerator approach may lower the
initial set-up costs and the operational costs, because the number of server boards, network switches, the
rack space requirements and power consumption may be lower than with a CPU-only approach.  Power consumption
estimates are beyond the scope of this work.  The same applies to equipment costs; however, prices are
available in the public domain and via commercial quotes.
\end{enumerate}

One should also note that the relative momentum resolution of the tracks obtained using the Hough transform is
not yet as good as that obtained using current HEP tracking~\cite{CMS:2010wta}. In order to match the curent
resolution, one would need to approximately double the time performance results obtained in
Figure~\ref{fig:TimePerformance}.

\section{Summary}

{\em Parallel} processing provides great promise for enhancing the performance of the trigger at the LHC and
extending its physics program. However, to achieve this goal, significant rethinking of the hardware and
algorithms have to be considered. These decisions impact initial acquisition and setup costs, rack space,
power limitations, operational costs associated with running the hardware and maintaining the software,
etc. In this article, the authors optimized a real time tracking algorithm based on Hough transform algorithm
on four different architectures Tesla GPU, Xeon and i7 CPUs, and \xphi\ coprocessor.  The outcome provides a
benchmark that may be used to identify the optimal computing system configuration for the required throughput,
cost and other compliance factors of the Hough transform based tracking detector.  These results suggest that
using an optimized algorithm on hybrid systems with add on coprocessors could be the leap necessary to
discover new physics at the LHC.

\bibliographystyle{elsarticle-num}
\bibliography{HTMPPE}

\end{document}